\documentstyle[12pt]{article}

\begin{document}

\author{L. Herrera
\thanks{\'Area de F\'\i sica Te\'orica, 
Facultad de Ciencias, Universidad de Salamanca, 37008 Salamanca,
Espa\~na.}
\thanks{Permanent address: Departamento de F\'\i sica,
        Facultad de Ciencias, Universidad Central de Venezuela, 
        Caracas, Venezuela and Centro de Astrof\'\i sica Te\'orica 
        (C.A.T.), M\'erida, Venezuela.} 
and J. Mart\'\i nez \thanks{Grupo de F\'\i sica Estad\'\i stica,
Departamento de F\'\i sica,
Universidad Aut\'onoma de Barcelona, 08193 Bellaterra, Barcelona,
Espa\~na. e-mail:justino@telemaco.uab.es, phone:+34-3-581 25 75}}
\title{Gravitational collapse: A case for thermal relaxation}
\date{}
\maketitle

\begin{abstract}
Two relativistic models for collapsing spheres at different stages of
evolution, which include pre-relaxation processes, are presented. The
influence of relaxation time on the outcome of evolution in both cases is
exhibited and established. It is shown that relaxation
processes can drastically change the final state of the collapsing system.
In particular, there are cases in which the value of the relaxation time
determines the bounce or the collapse of the sphere.
\end{abstract}
\newpage
\section{Introduction}

In the study of gravitational collapse, where dissipative processes play a
fundamental role, thermal relaxation time ($\tau$) is usually neglected. The
reason for this may be found in the well known fact that, for most materials
at usual laboratory conditions, $\tau$ is very small as compared with
typical scales of time of most self-gravitating systems. Thus, it is of the
order of $10^{-11}$ s for phonon-electron interaction and of the order of $%
10^{-13}$ s for phonon-phonon and free electron interaction at room
temperature \cite{Peirls}.

There are however situations where relaxation time may not be negligible.
Thus, for example, for superfluid Helium, $\tau$ is of the order of $10^{-3}
$ s for a temperature $1.2$ K \cite{BaMe48}.

Also, and much more important in the context of the problem considered here,
in cores of evolved stars the electron gas is highly degenerate. Then, since
the quantum cells of phase space are filled up, such that collisions in
which the momentum is changed become rather improbable, the mean free path
increase considerably, increasing $\tau$ thereby.

Thus, for example, for a completely degenerate core of radius $\sim 10^{-2}
R_\odot$ at $T \sim 10^7$ K, the thermal relaxation time is of the order of
one second \cite{Ha88}. However, the order of magnitude of relaxation time
used here is much lower than that.

Using the expression for the thermal conductivity by electrons in neutron
star matter introduced by Flowers and Itho \cite{FlIt79a,FlIt79b},
we found (see
below) without difficulties relaxation times in the range $\left[10^{-6}
\mbox{ s},10^{-4} \mbox{ s}\right]$.

There have been recent calculations on collapsing systems which include
pre-relaxation processes \cite{PrHeEs96,Martinez96}. However, either they
consider a constant value for the conductivity \cite{PrHeEs96} or they are
calculated for a single value of $\tau$ \cite{Martinez96}, excluding the
possibility of comparing the evolution for different relaxation times, and
thereby, of assessing the influence of $\tau$ on the evolution of the
collapsing object.

In this paper we describe the evolution of two different self-gravitating
systems dissipating energy through a radial heat flow vector.

Modeling is achieved by using the HJR formalism 
\cite{HeJiRu80,CoHeEsWi82,HeJiEs87}. However, unlike Di Prisco, Herrera
and Esculpi \cite{PrHeEs96} we shall not
consider a constant value for thermal conductivity and the astrophysical
setting is much more realistic. Also, unlike Mart\'\i nez
\cite{Martinez96} we shall
follow the approach introduced by Di Prisco et al
\cite{PrHeEs96} which allows for comparing
the evolution for different values of $\tau$.

The first model is based on the well known Tolman VI solution \cite{Tolman39}
and may be accomodated to describe to some extent, the core implosion and
subsequent bounce, at earlier stages of a supernova explosion.

The second one is more suitable for describing the Kelvin-Helmholtz phase of
the birth of a neutron star \cite{BuLa86}. However, it is important to
emphasize that in spite of the fact that the order of magnitude of different
physical variables are well within the range of expected values, our main
goal here is not to present a detailed modeling of those scenarios, but to
bring out the relevance of pre-relaxation processes in situations when
degenerate cores are involved.

The paper is organized as follows. In the next section the field equations
and conventions are presented and also a brief resume of the HJR formalism
is given. The heat conduction equation is presented in section 3, as well
as the approach to operate the HJR formalism.

Section 4 is devoted to the description of the two models considered in this
paper.

Finally, some conclusions and comments are included in the last section.

Partial differentiation with respect to $u$ and $r$ are indicated by means
of the subscripts 0 and 1 respectively. The subscript $a$ denote that the
quantity is evaluated at the surface.

\section{The HJR formalism}

We shall consider a non-static distribution of matter which is spherically
symmetric, and consists of fluid (which may be locally anisotropic) and heat
flow. In Bondi coordinates \cite{BoBuMe62,Bondi64}, the metric takes the
form
\begin{equation}
\label{Bondi}ds^2=e^{2\beta }\left[ \frac Vrdu^2+2dudr\right] -r^2\left(
d\theta ^2+sin^2\theta d\phi ^2\right) ,
\end{equation}
where $u~=~x^0$ is a time like coordinate ($g_{uu}~>~0$) , $r~=~x^1$ is the
null coordinate ($g_{rr}~=~0$) and $\theta ~=~x^2$ and $\phi ~=~x^3$ are the
usual angle coordinates. The $u$-coordinate corresponds to the retarded time
in flat space-time and the metric functions $\beta $ and $V$ are functions
of $u$ and $r$. A function $\tilde m(u,r)$ can be defined by
\begin{equation}
V=e^{2\beta }(r-2\tilde m(u,r)),
\end{equation}
which is the generalization, inside the distribution, of the ``mass aspect''
defined by Bondi et al \cite{BoBuMe62}. In the static limit it coincides
with the Schwarzschild mass. On the other hand, the radiation coordinates $u$%
, $r$, $\theta $ and $\phi $ can be related to Schwarzschild ones ($%
T,R,\Theta ,\Phi $) by
\begin{equation}
T=u+\int_0^r\frac rVdr,
\end{equation}
\begin{equation}
R=r,\ \ \ \Theta =\theta ,\ \ \ \Phi =\phi ,
\end{equation}
and to local Minkowskian coordinates ($t$, $x$, $y$, $z$) by
\begin{equation}
\label{ccoor}dt=e^\beta (\sqrt{\frac Vr}du+\sqrt{\frac rV}dr),
\end{equation}
\begin{equation}
\label{xccor}dx=e^\beta \sqrt{\frac rV}dr,
\end{equation}
\begin{equation}
\label{ycoor}dy=rd\theta ,
\end{equation}
\begin{equation}
\label{zcoor}dz=rsin\theta d\phi .
\end{equation}

At the outside of the fluid distribution the space-time is described by the
Vaidya metric \cite{Vaidya51}, which in Bondi coordinates is given by $\beta
=0$ and $V=r-2m(u).$

For the matter distribution considered here, the stress-energy tensor can be
written as \cite{Martinez96}
\begin{equation}
T_{\mu \nu }=(\rho +P_{\perp })U_\mu U_\nu -P_{\perp }g_{\mu \nu
}+(P_r-P_{\perp })\chi _\mu \chi _\nu +2{\cal Q}_{(\mu }U_{\nu )},
\end{equation}
where $\rho $, $P_r$, $P_{\perp }$ are the energy density, radial pressure
and tangential pressure respectively as measured by a Minkowskian observer
in the Lagrangean frame. The stress-energy tensor outlined above is obtained
as measured by an observer using Bondi coordinates. For this observer the
four-velocity $U_\mu $, and the heat flux vector ${\cal Q}^\mu $ are given
by
\begin{equation}
\label{umu}U_\mu =e^\beta \left( \sqrt{\frac Vr}\frac 1{(1-\omega ^2)^{1/2}},%
\sqrt{\frac rV}\left( \frac{1-\omega }{1+\omega }\right) ^{1/2},0,0\right) ,
\end{equation}
and
\begin{equation}
\label{qmu}{\cal Q}^\mu ={\cal Q}e^{-\beta }\left( -\sqrt{\frac rV}\left(
\frac{1-\omega }{1+\omega }\right) ^{1/2},\sqrt{\frac Vr}\frac 1{(1-\omega
^2)^{1/2}},0,0\right) ,
\end{equation}
whereas $\chi _\mu =-{\cal Q}_\mu /{\cal Q}$, ${\cal Q}=\sqrt{-{\cal Q}^\mu
{\cal Q}_\mu }$ is the heat flow, and $\omega$ is the velocity of matter as
measured by the Minkowski observer defined by (\ref{ccoor})--(\ref{zcoor}).

The Einstein field equations, inside the matter distribution, can be written
as \cite{HeJiEs87}:

\begin{equation}
\label{ecu00}\frac 1{4\pi r(r-2\tilde m)}\left( -\tilde m_0e^{-2\beta }+(1-2%
\tilde m/r)\tilde m_1\right) =\frac 1{1-\omega ^2}(\rho +2\omega {\cal Q}%
+P_r\omega ^2),
\end{equation}
\begin{equation}
\label{ecu01}\frac{\tilde m_1}{4\pi r^2}=\frac 1{1+\omega }(\rho -{\cal Q}%
(1-\omega )-P_r\omega ),
\end{equation}
\begin{equation}
\label{ecu11}\beta _1\frac{r-2\tilde m}{2\pi r^2}=\frac{1-\omega }{1+\omega }%
(\rho -2{\cal Q}+P_r),
\end{equation}
\begin{equation}
\label{ecu22}-\frac{\beta _{01}e^{-2\beta }}{4\pi }+\frac 1{8\pi }(1-2\frac{%
\tilde m}r)(2\beta _{11}+4\beta _1^2-\frac{\beta _1}r)+\frac{3\beta _1(1-2%
\tilde m_1)-\tilde m_{11}}{8\pi r}=P_{\perp },
\end{equation}
while outside matter, the stress-energy tensor takes the form
\begin{equation}
T_{\mu \nu }=-\frac 1{4\pi r^2}\tilde m_0\delta _\mu ^u\delta _\nu ^u
\end{equation}
and the only non trivial Einstein equation reads
\begin{equation}
\widetilde{m}_0=-4\pi r^2\varepsilon \left( 1-\frac{2\widetilde{m}(u)}r%
\right) .
\end{equation}
where $\varepsilon $ is proportional to the energy density of the radiation
traveling in the radial direction -- see Bondi \cite{Bondi64} for
details, and subscripts 0,1
denote derivatives with respect to $u$ and $r$ respectively.

In the HJR formalism \cite{HeJiRu80,CoHeEsWi82,HeJiEs87}, one introduces the
concept of effective energy density and effective pressure. From (\ref{ecu01}%
), the mass function can be expressed as
\begin{equation}
\label{masa}\tilde m=\int_0^r4\pi r^2\tilde \rho dr,
\end{equation}
where
\begin{equation}
\label{rotilde}\tilde \rho =\frac 1{1+\omega }(\rho -{\cal Q}(1-\omega
)-P_r\omega ),
\end{equation}
is the effective energy density, which in the static limit reduces to the
energy density of the system.

In a similar way, we can define the effective pressure. From (\ref{ecu11})
one has
\begin{equation}
\beta =\int_{a(u)}^r\frac{2\pi r^2}{r-2\tilde m}\frac{1-\omega }{1+\omega }%
(\rho -2{\cal Q}+P_r)dr,
\end{equation}
or
\begin{equation}
\label{betatil}\beta =\int_{a(u)}^r\frac{2\pi r^2}{r-2\tilde m}(\tilde \rho +%
\tilde P)dr,
\end{equation}
with
\begin{equation}
\label{ptilde}\tilde P=\frac 1{1+\omega }(-\omega \rho -{\cal Q}(1-\omega
)+P_r)
\end{equation}
being the effective pressure. This one, as the effective energy density,
only has a clear physical meaning in the static case, in which it reduces to
the radial pressure.

Matching the Vaidya metric to the Bondi metric at the surface ($r=a$) of the
fluid distribution implies $\beta _a=\beta(u,r=a)=0$ and the continuity of
the mass function $\tilde m(u,r)$. The continuity of the second fundamental
form must be demanded as well, leading to condition
\begin{equation}
\label{jc3}\dot a=-\left( 1-2\frac{\tilde m_a}a\right) \frac{\tilde P_a}{%
\tilde P_a+\tilde \rho _a}.
\end{equation}
-- see Herrera and Jim\'enez \cite{HeJi83} for details -- where
overdot denotes derivative with
respect to $u$.

The well-known condition
\begin{equation}
\label{jc6}{\cal Q}_a=P_{ra},
\end{equation}
for radiative spheres \cite{Santos85} can be obtained from the coordinate
transformation (\ref{ccoor}). Effectively, the velocity of matter in Bondi
coordinates can be written as
\begin{equation}
\label{drdu}\frac{dr}{du}=\frac Vr\frac \omega {1-\omega },
\end{equation}
evaluating the last expression at the surface and comparing it with (\ref
{jc3}) it follows
\begin{equation}
\label{jc5}\tilde P_a=-\omega _a\tilde \rho _a.
\end{equation}
It is easy to show, using (\ref{rotilde}) and (\ref{ptilde}), that this
condition is equivalent to (\ref{jc6}).

The HJR method is based in a system of three surface equations which allows
us to find the $u$ dependence of the functions $\beta$ and $\tilde{m}$
present in the field equations (\ref{ecu00}-\ref{ecu22}).

To derive the surface equations, usually five dimensionless variables are
defined
\begin{equation}
\label{adime}A\equiv \frac a{\widetilde{m}(0)}\;\;\;\;M\equiv \frac{%
\widetilde{m}}{\widetilde{m}(0)}\;\;\;\;u\equiv \frac u{\widetilde{m}(0)}%
\;\;\;\;F\equiv 1-\frac{2M}A\;\;\;\;\Omega \equiv \frac 1{1-\omega _a},
\end{equation}
where $\widetilde{m}(0)$ is the initial total mass of the system. Then,
using (\ref{adime}) and (\ref{jc5}), (\ref{jc3}) yields the first surface
equation. Using the functions just defined into (\ref{jc3}) we get the first
surface equation
\begin{equation}
\label{se1}\dot A=F(\Omega -1),
\end{equation}
which gives the evolution of the radius of the star. From now on, a dot
denotes derivative with respect to the dimensionless variable $u/\widetilde{m%
}(0)$.

The second surface equation can be obtained from the luminosity evaluated at
the surface of the system. The luminosity as measured by an observer at rest
at infinity reads
\begin{equation}
\label{luminosity}L=-\dot M=\frac E{(1+z_a)^2}=EF=\hat E(2\Omega -1)F=4\pi
A^2{\cal Q}_a\left( 2\Omega -1\right) F,
\end{equation}
where $z_a$ refers to the boundary gravitational redshift, $\hat E$ is the
luminosity as seen by a comoving observer, and $E$ is the luminosity
measured by a non comoving observer located on the surface. Using
relationship (\ref{ecu00}) and (\ref{luminosity}) together with the first
surface equation we obtain the second one as
\begin{equation}
\label{se2}\dot F=\frac{2L+F(1-F)(\Omega -1)}A,
\end{equation}
which expresses the evolution of the redshift at the surface.

The third surface equation is model dependent. For anisotropic fluids the
relationship $(T_{r;\mu }^\mu )_a=0$ can be written as
\[
\frac{\dot F}F+\frac{\dot \Omega }\Omega -\frac{\stackrel{.}{\tilde \rho }_a%
}{\tilde \rho _a}+F\Omega ^2\frac{\tilde R_{\perp _a}}{\tilde \rho _a}-\frac
2AF\Omega \frac{P_{ra}}{\tilde \rho _a}=
\]
\begin{equation}
\label{se3}(1-\Omega )\left[ 4\pi A\tilde \rho _a\frac{3\Omega -1}\Omega -%
\frac{3+F}{2A}+F\Omega \frac{\tilde \rho _{1_a}}{\tilde \rho _a}+\frac{%
2F\Omega }{A\tilde \rho _a}(P_{\perp }-P_r)_a\right] ,
\end{equation}
being
\begin{equation}
\tilde R_{\perp _a}=\tilde P_{1_a}+\left( \frac{\tilde P+\tilde \rho }{1-2%
\tilde m/r}\right) _a\left( 4\pi r\tilde P+\frac{\tilde m}{r^2}\right)
_a-\left( \frac 2r(P_{\perp }-P_r)\right) _a.
\end{equation}
Expression (\ref{se3}) generalizes the Tolman-Oppenheimer-Volkov equation to
the non-static radiative anisotropic case.

The HJR method \cite{HeJiRu80} allows us to find non static solutions of the
Einstein equations from static ones. The algorithm, extended for anisotropic
fluids, can be summarized as follows \cite{HeJiRu80,CoHeEsWi82}

\begin{enumerate}
\item  Take a static but otherwise arbitrary interior solution of the
Einstein equations for a spherically symmetric fluid distribution
\begin{equation}
P_{st}=P(r),\ \ \ \ \ \rho _{st}=\rho (r).
\end{equation}

\item  The effective quantities $\tilde \rho \equiv \tilde \rho (u,r)$ and $%
\tilde P\equiv \tilde P(u,r)$ must coincide with $\rho _{st}$ and $P_{st}$
respectively in the static limit. We assume that the $r$-dependence in
effective quantities is the same that in its corresponding static ones.
Nevertheless, note that junction conditions in terms of effective variables,
read as (\ref{jc5}). This condition allows us to find out the relation
between the $u$-dependence of $\tilde \rho \equiv \tilde \rho (u,r)$ and $%
\tilde P\equiv \tilde P(u,r)$.

The rationale behind the assumption on the $r$ dependence of the{\it \
effective variables }$\tilde{P}$ and $\tilde{\rho}$, can be grasped in terms
of the characteristic times for different processes involved in a collapse
scenario. If the hydrostatic time scale ${\cal T}_{HYDR}$, which is of the
order $\sim 1/\sqrt{G\rho }$ (where $G$ is the gravitational constant and $%
\rho $ denotes the mean density) is much smaller than the {\it %
Kelvin-Helmholtz} time scale ( ${\cal T}_{KH}$ ), then in a first
approximation the inertial terms in the equation of motion can be ignored.
 Therefore in this first approximation (quasi-stationary approximation) the $r$
dependence of $P$ and $\rho $ are the same as in the static solution. Then
the assumption that the{\it \ effective variables}  have the same $r$
dependence as the physical variables of the
static situation, represents a correction to that approximation.

\item  Introduce $\tilde \rho (u,r)$ and $\tilde P(u,r),$ into (\ref{masa})
and (\ref{betatil}) to determine $\tilde m$ and $\beta $ up to three unknown
functions of time.

\item  The three surface equations form a system of first order ordinary
differential equations, by solving it we find the evolution of the radius, $%
A(u)$, and two unknown functions of time. These ones can be related with the
$u$-dependence of $\tilde \rho \equiv \tilde \rho (u,r)$ and $\tilde P\equiv
\tilde P(u,r)$.

\item  There are four unknown functions of time ($A$, $F$, $\Omega $ and $L$%
). Thus, it is necessary to impose the evolution of one of them to solve the
system of three surface equations. Usually the luminosity is taken as an
input function because it can be found from observational data.

\item  Once these three functions are known, it is easy to find $\tilde m$
and $\beta .$ Therefore, the interior metric is completely defined.

\item  Now, the left-hand side of the Einstein equations (\ref{ecu00})-(\ref
{ecu22}) is known. However, the right-hand side of these equations contain
five unknown quantities ($\omega $, $\rho $, $P_r$, $P_{\perp }$ and ${\cal Q%
}$). Thus, it is necessary to supply another equation to close the system of
field equations. In the anisotropic static case the equation of hydrostatic
equilibrium reads \cite{CoHeEs81},
\begin{equation}
P_{\perp }-P_r=\frac r2P_1+\left( \frac{\rho +P}2\right) \left( \frac{m+4\pi
r^3P}{r-2m}\right) .
\end{equation}
This expression is usually generalized, in the context of HJR method, to
non-static cases by substituting the physical quantities by the effective
variables \cite{CoHeEsWi82,BaRo92}
\begin{equation}
\label{ptmpr}P_{\perp }-P_r=\frac r2\widetilde{P}_1+\left( \frac{\widetilde{%
\rho }+\widetilde{P}}2\right) \left( \frac{\widetilde{m}+4\pi r^3\widetilde{P%
}}{r-2\widetilde{m}}\right) .
\end{equation}
Now, the Einstein equations, supplemented with (\ref{ptmpr}), form a closed
system of equations, and quantities $\omega $, $\rho $, $P_r$, $P_{\perp }$
and ${\cal Q}$ can be found.
\end{enumerate}

\section{System of Surface Equations: Heat conduction equation.}

As we mentioned in the previous section, equations (\ref{se1}), (\ref{se2}),
and (\ref{se3}) constitute a system of surface equations. However, if we
desire to study the influence of the relaxation processes during the
collapse, it is necessary to introduce an hyperbolic transport equation.
Recently, Di Prisco, Herrera and Esculpi \cite{PrHeEs96} have shown the
importance of the relaxation processes involving the heat flow using a
Schwarzschild based model with constant thermal conductivity. Our aim is to
apply the same procedure to different and more realistic models in order to
discern the effects that are model independent and how sensitive is the
final state to the presence of relaxation processes.

The system of surface equations can be solved for a given luminosity
profile. Thus, the temperature could be found if $\tau$ were known \cite
{Martinez96}. Nevertheless, this procedure does not give us information
about the influence of the relaxation processes on the luminosity profile.
Because of this, it seems more convenient to our purposes to follow the
method outlined by Di Prisco et al \cite{PrHeEs96}.

First of all, we assume the evolution of the heat flow governed by the
Maxwell-Cattaneo transport equation,
\begin{equation}
\label{MC}\tau \frac{d{\cal Q}^\nu}{ds}h_\nu ^\mu +{\cal Q}^\mu =\chi h^{\mu
\nu }\left[ T_{,\nu }-T\frac{d{U}_\nu}{ds} \right] ,
\end{equation}
where
\[
\frac{d{\cal Q}^\nu}{ds}=U^\mu{\cal Q}^{\nu}_{;\mu},
\]
\[
\frac{dU^\nu}{ds}=U^\mu U^{\nu}_{;\mu},
\]
and $\chi $ is the thermal conductivity coefficient. The non vanishing
covariant components of the four acceleration read:
\[
\frac{d{U}_r}{ds}=\frac 1{1+\omega }\left[ \frac 1{2r}-2\beta _1-\frac{1-2%
\widetilde{m}_1}{2(r-2\widetilde{m})}\right] +re^{-2\beta }\left( \frac{%
1-\omega }{1+\omega }\right) \frac{\widetilde{m}_0}{(r-2\widetilde{m})^2}
\]
\begin{equation}
\label{upuntor}-\frac 1{(1+\omega )^2(1-\omega )}\left[ \omega \omega
_1+re^{-2\beta }\frac{1-\omega }{r-2\widetilde{m}}\omega _0\right] ,
\end{equation}
and
\begin{equation}
\label{upuntou}\frac{d{U}_u}{ds}=e^{{-2\beta }}\left( 1-\frac{2\tilde m}r%
\right) \frac \omega {1-\omega }\frac{d{U}_r}{ds}.
\end{equation}
It is easy to demonstrate that in our case, eq.(\ref{MC}) has only one
independent component. Thus, evaluating one of them at the surface we obtain
\[
\tau \stackrel{.}{\cal Q}_a+{\cal Q}_a\sqrt{F(2\Omega -1)}=
\]

\begin{equation}
\label{MCA0}\chi _a\left[ \stackrel{.}{T_a}-T_{1a}F(2\Omega -1)-T_aF(2\Omega
-1)\left( \frac{(1-F)\Omega}{2AF(2\Omega-1)}+\frac L{F^2A(2\Omega -1)}+\frac{%
\stackrel{.}{\Omega }}{F(2\Omega -1)^2}\right) \right] ,
\end{equation}
where we have used (\ref{upuntor}), (\ref{umu}), (\ref{qmu}), and (\ref
{Bondi}).

We shall assume that the thermal conduction is dominated by electrons. Thus,
the thermal conductivity is given by expression \cite{FlIt79a,FlIt79b}
\begin{equation}
\label{xi}\chi \simeq 10^{23}\frac{\rho _{14}}{T_8}\mbox{ erg. s}^{-1}%
\mbox{
cm}^{-1}\mbox{ K}^{-1},
\end{equation}
where the energy density and the temperature are given in $10^{14}$ g cm$%
^{-3}$ and $10^8$ K units respectively. In the HJR formalism the initial
mass is normalized to unity (\ref{adime}). Therefore, all quantities that,
in geometrized units, have dimensions of a power of length, are within this
framework dimensionless. The expression (\ref{xi}) for the thermal
conductivity coefficient can be rewritten as
\begin{equation}
\label{xiad}\chi \simeq \frac{{\cal C}}\xi \frac \rho T,
\end{equation}
being ${\cal C}=2.5126\times 10^{-20}$, $\rho $ the dimensionless energy
density, and $T$ the temperature in Kelvin. The initial mass has been
written in terms of the solar mass as $M_o=\xi M_{\odot }$, where $\xi$ is a
numerical factor.

The energy density in the surface can be written in terms of the heat flow
and the effective energy by means of the definition (\ref{rotilde}).
Evaluating it at the surface and using (\ref{adime}) and junction condition (%
\ref{jc6}) we obtain,
\begin{equation}
\label{rhoa}\rho _a={\cal Q}_a+\widetilde{\rho }_a\left[ \frac{2\Omega -1}%
\Omega \right] .
\end{equation}
Then, evaluating (\ref{xiad}) at the surface, the heat transport equation at
the surface (\ref{MCA0}) reads%
\[
\tau \stackrel{.}{\cal Q}_a+{\cal Q}_a\sqrt{F(2\Omega -1)}=\frac{{\cal C}}\xi
\left( {\cal Q}_a+\widetilde{\rho }_a\left[ \frac{2\Omega -1}\Omega \right]
\right) \times
\]
\begin{equation}
\label{MCA}\left[ \frac{\stackrel{.}{T_a}}{T_a}-\frac{T_{1a}}{T_a}F(2\Omega
-1)-F(2\Omega -1)\left( \left( \frac{1-F}{2AF}\right) \frac \Omega {2\Omega
-1}+\frac L{F^2A(2\Omega -1)}+\frac{\stackrel{.}{\Omega }}{F(2\Omega -1)^2}%
\right) \right] .
\end{equation}

\subsection{Boundary condition for the transport equation}

Expression (\ref{MCA}) relates the heat flow and the temperature at the
surface to the temperature gradient through quantities present in the
surface equations (\ref{se1}), (\ref{se2}), and (\ref{se3}).

The connection, at the surface, between the heat flow and the temperature
can be found by means of the effective temperature $T_{eff}$. This one is
defined by means of the luminosity perceived by an observer located
momentarily on the surface,
\begin{equation}
\label{eteff}E=\left[ 4\pi r^2\sigma T_{eff}^4\right] _{r=a},
\end{equation}
being $\sigma $ the Steffan-Boltzmann radiation constant. Thus, the
effective temperature would be the temperature at the surface of the star if
it would radiate as a black body. This concept is linked up with the theory
of stellar atmospheres by means of the photosphere \cite[p. 70]{KiWe94},
\cite[p. 586]{ShTe83} and \cite[p. 295]{HaKa94}. The photosphere comprises
the most external layers of the star, and its thickness is determined by the
optical depth for photons. At the surface this optical depth vanishes and
the effective temperature is related to the material one by the expression
\begin{equation}
\label{teff}T_a^4=\frac 12T_{eff}^4.
\end{equation}

Then, by substitution of (\ref{teff}) and (\ref{eteff}) into (\ref
{luminosity}) we obtain
\begin{equation}
{\cal Q}_a=T_a^4\frac{2\sigma }{2\Omega -1}.
\end{equation}
Introducing this expression in (\ref{MCA}), we obtain the fourth surface
equation
\[
\stackrel{.}{y}\left[ \frac{2y^4\xi ^2}{2\Omega -1}\left( 4\tau -\frac{{\cal %
C}}\xi \right) -\widetilde{\rho }_a\frac{{\cal C}}\xi \left( \frac{2\Omega -1%
}\Omega \right) \right] =\xi ^2y^5\left[ \frac{4\tau \stackrel{.}{\Omega }}{%
(2\Omega -1)^2}-\frac{2\sqrt{F}}{\sqrt{2\Omega -1}}\right] -
\]
\begin{equation}
\label{se4}\frac{{\cal C}}\xi \left[ \frac{2y^4\xi ^2}{2\Omega -1}+%
\widetilde{\rho }_a\left( \frac{2\Omega -1}\Omega \right) \right] \left(
y_1F(2\Omega -1)+y\Phi \right) ,
\end{equation}
where
\begin{equation}
\label{fi}\Phi =\left( \frac{1-F}{2A}\right) \Omega +8\pi Ay^4\xi ^2+\frac{%
\stackrel{.}{\Omega }}{2\Omega -1}.
\end{equation}
the function $y$ has been defined as
\begin{equation}
y^4\equiv \zeta T_a^4,
\end{equation}
and the constant $\zeta \equiv \sigma M_{\odot }^2\simeq 3.4097\times
10^{-54}$ K$^{-4}$ is calculated taken $M_{\odot }$ in geometrized units.
The luminosity $L$, present in (\ref{MCA}), can be written in terms of $y$
by means of (\ref{luminosity}), (\ref{teff}) and (\ref{eteff})
\begin{equation}
\label{lumi}L=8\pi A^2y^4\xi ^2F.
\end{equation}
Note that all terms in (\ref{se4}) are dimensionless, including the
relaxation time $\tau $, which is related to the physical $\tau (\equiv \tau
_{ph})$ by means of
\begin{equation}
\tau =\tau _{ph}\frac{c^3}{Gm_o}\simeq 2.0298\times 10^5\frac{\tau _{ph}}\xi
\mbox{ s}^{-1}.
\end{equation}
where $c$ and $G$ denotes the speed of light and the gravitational constant,
and $m_o$ is the total mass in grams.

\subsection{The DHE approach.}

As we mentioned above we shall use the approach introduced by Di
Prisco et al \cite{PrHeEs96} (DHE approach henceforth) to solve the
system of surface equations. This system of equations usually is composed of
three differential equations which give the evolution of three quantities:
The radius (\ref{se1}), the boundary gravitational redshift (\ref{se2}) and
the collapse velocity of the surface (\ref{se3}). In order to study the
effect of the thermal relaxation processes on the system it is necessary to
include an additional equation accounting for the evolution of the
temperature at the surface. Thus, expression (\ref{se4}) constitutes the
fourth surface equation. The method can be described as follows:

\begin{enumerate}
\item  Solve, as usual, the system of surface equations (\ref{se1}), (\ref
{se2}), and (\ref{se3}) for a given luminosity $L.$

\item  Introduce the results into the fourth surface equation (\ref{se4}),
and solve if for $\tau =0$ (Eckart's law). Since $L$ is known, equation (\ref
{lumi}) allows us to find the temperature evolution and equation (\ref{se4})
provides us the temperature gradient.

\item  Use the evolution found for $T_{1a}$ in the resolution of the
complete system of surface equations (\ref{se1}), (\ref{se2}), (\ref{se3}),
and (\ref{se4}) for non vanishing $\tau $ (Maxwell-Cattaneo law)$.$ So that
the evolution of $A,F$ and $L$ is found for a given value of $\tau $ and we
are able to compare it with that found for $\tau =0.$
\end{enumerate}

The method described above must be self consistent, {\it i.e.} we must
recover the luminosity profile assumed in the first point if a vanishing
relaxation time is imposed in the third one.

The procedure adopted here assumes that the temperature gradient is
independent of the adopted value for $\tau $. This hypothesis can be
justified in the following terms: The existence of a temperature gradient is
caused by local processes. As a consequence of its presence a non-vanishing
flux appears, giving rise to dissipative processes. The relation between
cause and effect is, in this case, non linear due to the fact that
dissipative processes can change the local situation. On the other hand,
relaxation processes act directly on dissipative processes. Thus, they must
affect in a indirectly way to the local situation. Nevertheless, the effect
on local processes must be less important than on dissipative ones having,
then, a little influence on temperature gradients. In first approximation we
assume that, for time-scales of the order of $\tau$, the influence of the
relaxation processes on temperature gradients can be neglected.

In order to ascertain the goodness of this assumption, we have numerically
integrated the system of surface equations (\ref{se1}), (\ref{se2}), (\ref
{se3}) and (\ref{se4}), for the models presented below, assuming the same
luminosity profile for a wide range of values of $\tau $. The differences
found among temperature gradients for different $\tau $ can be neglected.
This fact reinforces the adopted hypothesis. The unsensibility of the inner
temperature to the relaxation time for a fixed luminosity \cite{Martinez96}
points in the same direction as well.

\section{The models}

The HJR method starts from a static solution of the Einstein equations (sec.
II). We shall study the evolution of two different models. The first one,
corresponds to the well-known Tolman VI solution \cite{Tolman39}, while the
second one comes from a static solution due to Gokhroo and Mehra \cite
{GoMe94}.

\subsection{Tolman VI-type solution}

The equation of state obtained from the Tolman VI static solution
approaches, in the core of the star, to a highly compressed Fermi gas. The
energy density and radial pressure are
\begin{equation}
\label{dentol}\rho(r)_{st} =\frac 3{56\pi r^2},
\end{equation}
and
\begin{equation}
\label{prestol}P(r)_{st}=\frac \rho 3\left[ \frac{9a-9r}{9a-r}\right]
\end{equation}
respectively. In order to generate non static solutions, Tolman VI model has
been used often in the HJR framework as a departing solution \cite{HeJiRu80}%
. Following the HJR method, the effective quantities are taken as
\begin{equation}
\label{dentoltil}\widetilde{\rho }=\frac{3G(u)}{r^2},
\end{equation}
and
\begin{equation}
\label{ptoltil}\widetilde{P}=\frac{\widetilde{\rho }}3\left[ \frac{1-9K(u)r}{%
1-K(u)r}\right] .
\end{equation}
Using expressions (\ref{dentoltil}), (\ref{masa}) and (\ref{adime}) the
function $G(u)$ reads,
\begin{equation}
\label{gtol}G=\frac{1-F}{24\pi }.
\end{equation}
Function $K(u)$ follows, as a function of $\Omega $, from junction condition
(\ref{jc5}), (\ref{dentoltil}), (\ref{ptoltil}) and (\ref{adime}),
\begin{equation}
\label{ktol}K=\frac{1}{3A}\left[ \frac{3-4\Omega }{1-4\Omega }\right] .
\end{equation}

By means of the last four expressions, the system of surface equations (\ref
{se1}), (\ref{se2}), (\ref{se3}), and (\ref{se4}) reads
\begin{equation}
\label{tol1}\dot A=F(\Omega -1),
\end{equation}
\begin{equation}
\label{tol2}\dot F=\frac{2L+F(1-F)(\Omega -1)}A,
\end{equation}
\begin{equation}
\label{tol3}\frac{\dot F}F+\left( 1-F\right) \frac{\dot \Omega }\Omega =%
\frac{F(1-F)(4\Omega -3)(4\Omega -1)}{8A}-\frac{(1-F)^2}{2A\Omega }+\frac{%
4L\Omega }{(2\Omega -1)A},
\end{equation}
and%
\[
\stackrel{.}{y}\left[ \frac{2y^4\xi ^2}{2\Omega -1}\left( 4\tau -\frac{{\cal %
C}}\xi \right) -\frac{{\cal C}}\xi \left( \frac{1-F}{8\pi A^2}\right) \left(
\frac{2\Omega -1}\Omega \right) \right] =\xi ^2y^5\left[ \frac{4\tau
\stackrel{.}{\Omega }}{(2\Omega -1)^2}-\frac{2\sqrt{F}}{\sqrt{2\Omega -1}}%
\right] -
\]
\begin{equation}
\label{tol4}\frac{{\cal C}}\xi \left[ \frac{2y^4\xi ^2}{2\Omega -1}+\left(
\frac{1-F}{8\pi A^2}\right) \left( \frac{2\Omega -1}\Omega \right) \right]
\left( y_1F(2\Omega -1)+y\Phi \right) ,
\end{equation}

This model can be applied in the context of the bouncing of the core during
a supernova explosion. It is important to note that the value of the
quantity $M/A$ plays a fundamental role in the evolution of the system. In
particular, for the model under consideration the bounce of the surface
occurs if $M/A\leq 3/14$ \cite{HeJiRu80}. As Di Prisco et al \cite
{PrHeEs96} showed, this quantity is influenced by the inclusion of
relaxation processes. If this effect is model independent then, in some
cases, it will cause a drastic change in the final state of the collapsing
core.

In order to evaluate the influence of non vanishing $\tau$ in the evolution
of the system, we assume as initial configuration:
\begin{equation}
\label{inicontol}A_o=6,\; \Omega_o=0.878497,\; F=\frac23,\; \mbox{and }
M_o=1.3M_{\odot}.
\end{equation}
This corresponds to a core with an energy density about $10^{14}$ g cm$^{-3}$
in the external layers, whereas its radius is $11521$ m.

In the case of vanishing relaxation time, the luminosity is taken as
\begin{equation}
\label{lumitol}L=\frac{2M_r}{\sqrt{x^3\pi}}\sqrt{t}e^{-t/x},
\end{equation}
where $M_r$ is the total radiated mass, and $x$ is an arbitrary parameter
which determines the decaying time of $L$. In this case $x$ has been taken
as $10$ and $M_r=8.407\times 10^{-6}$. This brings a maximum luminosity
around $1.5\times 10^{53}$ erg s$^{-1}$, and a characteristic time for $L$
close to $0.5$ ms. The total radiated energy is of the order of $10^{49}$
erg.

Following the DHE method described in the previous section, the system of
surface equations (\ref{tol1})-(\ref{tol4}) can be solved for $\tau=0$ using
(\ref{lumitol}). This allows us to find the temperature gradient in the
surface, which is used to solve (\ref{tol1})-(\ref{tol4}) for $\tau \not = 0$%
.

The luminosity profile is displayed in fig. 1 for several values of $\tau $.
The maximum luminosity decreases if $\tau $ grows. Thus, the higher $\tau $,
the softer collapse. Other effect that can be inferred from figure 1 refers
to the width of the pulse of luminosity. This one is larger for larger $\tau
$'s. The same influence can be observed in the temperature of the surface
(fig. 2). Note that the maximum luminosity and temperature takes place at
larger times as $\tau $ increases. This is due, as corresponds to a
relaxation process, to the fact that for larger values of the relaxation
time, the system requires more time to establish the heat flow.

As we mentioned above, the quantity $M/A$ is greatly influenced by the
presence of relaxation processes (figure 3). An interesting effect must be
noted in its evolution. The $M/A$ profile presents a sort of bifurcation for
$\tau =\tau _{bif}\sim 3.5\times 10^{-6}$ s. If $\tau >\tau _{bif}$, the
core evolves towards a more compact final state than the initial one. The
situation is reversed for $\tau <\tau _{bif}$. Thus, small values of $\tau $
have a great influence in the final state. Something similar occurs in the
evolution of the radius (figure 4). The bounce of the surface takes place if
$\tau <\tau _{bif}$, whereas for $\tau >\tau _{bif}$ the collapse is
ensured. This fact can be explained taking into account the influence of the
luminosity in the evolution of the collapse. In the present model, and for $%
\tau =0$, the pulse of luminosity is the responsible of the bounce of the
surface. In case when luminosity is low, the explosion won't take place. The
presence of relaxation processes imposes a lower luminosity. Therefore, for $%
\tau >\tau _{bif}$ the luminosity will be insufficient to stop the collapse,
and the surface won't bounce. There are cases in which, as in the studied
model, small variations in the luminosity make the bounce impossible.

The influence of thermal relaxation processes seems to be important in the
evolution of the collapsing system. Nevertheless, the question is if the
value found for $\tau _{bif}$ is close enough to the expected value for the
relaxation time. Assuming that the thermal conduction is dominated by
electrons, the relaxation time can be roughly estimated, in the limit of
high frequencies, by the expression \cite{PrHeEs96}
\begin{equation}
\tau \sim \frac{10^{20}}{T^2v^2}\mbox{ s},
\end{equation}
where $T$ is measured in K and thermal signals travel with speed $v$ given
in cm s$^{-1}$. This expression is only valid assuming $\rho $ and $\chi $
as constants. However, it can be taken as an indicative of the order of
magnitude for $\tau $. Assuming $T\sim 10^{10}$ K, and that thermal signals
propagating approximately at $10^3$ cm s$^{-1}$ (the second sound speed in
superfluid helium) we obtain $\tau \sim 10^{-6}$ s. Thus, the obtained $\tau
_{bif}$ must be considered as a realistic value for $\tau $ and the effects
described above take a special importance.

The total radiated mass, as a function of $\tau $, for two different initial
and boundary conditions is presented in figure 5. One of them (TOL06
henceforth) corresponds to initial conditions (\ref{inicontol}) and boundary
condition (\ref{lumitol}). The other (TOL04 from now on) has been solved
using
\begin{equation}
A_o=6.5,\;\Omega _o=0.845215,\;F=\frac 9{13},\;\mbox{and }M_o=1.3M_{\odot },
\end{equation}
as initial conditions, and assuming a Gaussian pulse profile for the
luminosity for vanishing $\tau $. The total radiated mass, if $\tau =0$, is $%
M_r=10^{-4}M_o$, and the maximum luminosity is about $7.2\times 10^{54}$ erg
s$^{-1}$. The point marked A in figure 5 indicates the limit of validity for
TOL04 model. The Tolman VI solution fails for large times. In particular,
TOL04 model fails before the luminosity may vanish if $\tau $ is larger than
$\sim 0.07$ ms. TOL04 model is interesting because it clearly shows the non
linearity, between the total radiated mass, and $\tau $. This effect occurs
in TOL06 model as well, but is only visible for very small relaxation times
and the effect is not so accused. In spite of the maximum luminosity
decreasing as $\tau $ increases, the total time in which the system radiates
grows with the relaxation time, and the total radiated mass too.
Nevertheless, the maximum value of the total radiated mass is bounded. This
upper limit to $M_r$ is reached if relaxation time is of the order of the
width of the luminosity profile.

\subsection{Gokhroo \& Mehra-type solution}

This solution corresponds to an anisotropic fluid with variable energy
density. The choice of this solution is based on the fact, that it leads to
densities and pressures similar to the well-known Bethe-B\"orner-Sato (BBS)
{\it Newtonian} state equation \cite{ShTe83,Borner73,Demianski85}. Its
generalization to the HJR method has been done recently with good results
\cite{Martinez96}.

In this solution the energy density and radial pressure are assumed as \cite
{GoMe94}
\begin{equation}
\label{statenergy}\rho (r)_{st}=\rho _c\left( 1-k\frac{r^2}{a^2}\right)
\end{equation}
and
\begin{equation}
\label{statpressure}P_r(r)_{st}=P_c\left( 1-\frac{2m}r\right) \left( 1-\frac{%
r^2}{a^2}\right) ^n,
\end{equation}
where the constants $k$ and $n$ are within the range $0\leq k\leq 1$ and $%
n\geq 1.$ The central energy density $\rho _c,$ and central radial pressure $%
P_c$ are related by expression
\begin{equation}
\label{lambda}P_c=\lambda \rho _c.
\end{equation}
The tangential pressure is given by
\[
P_{\bot }(r)-P_r(r)=\frac r2\left[ P_r\right] _1+\left( \frac{\rho +P_r}2%
\right) \left( \frac{m(r)+4\pi r^3P_r}{r-2m(r)}\right)
\]
\begin{equation}
=\frac 3{10}\frac{kP_c}{a^2}\alpha r^4\left( 1-\frac{r^2}{a^2}\right) ^n+%
\frac{r^2}{2\left( 1-\frac{2m}r\right) }\Phi ,
\end{equation}
where
\[
\Phi =2P_c\left( 1-2\frac mr\right) ^2\left[ 2\pi P_c\left( 1-\frac{r^2}{a^2}%
\right) ^{2n}-\frac n{a^2}\left( 1-\frac{r^2}{a^2}\right) ^{n-1}\right]
\]
\begin{equation}
\label{phi}+\frac{\alpha \rho _c}2\left( 1-\frac{3k}5\frac{r^2}{a^2}\right)
\left( 1-k\frac{r^2}{a^2}\right) ,
\end{equation}
being
\begin{equation}
\label{alpha}\alpha =\frac{8\pi \rho _c}3.
\end{equation}

The effective quantities are defined, following the second point of the HJR
method, as -- see Mart\'\i nez \cite{Martinez96} for details --
\begin{equation}
\label{tilderho}\widetilde{\rho }=\widetilde{\rho }_c(u)\left( 1-K(u)\frac{%
r^2}{a^2}\right) ,
\end{equation}
and
\begin{equation}
\label{tildepres}\widetilde{P}=\widetilde{P}_c(u)\left( 1-2\frac{\widetilde{m%
}}r\right) \left( 1-G(u)\frac{r^2}{a^2}\right) ^n.
\end{equation}
Where the functions $\widetilde{\rho }_c(u)$ and $\widetilde{P}_c(u)$, are
defined by
\begin{equation}
\label{rhoc}\widetilde{\rho }_c(u)=\rho _c\frac{K(u)}{K(0)}\equiv \rho _c%
\frac K{K_o},
\end{equation}
and
\begin{equation}
\label{presc}\widetilde{P}_c(u)=P_c\frac{K(u)}{K(0)}\equiv P_c\frac K{K_o}.
\end{equation}
The functions of time $K(u)$, and $G(u)$ are found to be

\begin{equation}
\label{K}K(u)=\left\{
\begin{array}{cc}
\frac 56\left[ 1+\sqrt{1-\left( \frac{12K_o}{5\alpha }\right) \left( \frac{%
1-F}{A^2}\right) }\right] & \mbox{if }K_o>\frac 56 \\ \frac 56\left[ 1-\sqrt{%
1-\left( \frac{12K_o}{5\alpha }\right) \left( \frac{1-F}{A^2}\right) }%
\right] & \mbox{if }K_o<\frac 56
\end{array}
\right. ,
\end{equation}
and
\begin{equation}
\label{G}G(u)=1-\left[ \frac{\left( 1-\Omega \right) \left( 1-K\right) }{%
F\Omega \lambda }\right] ^{1/n}.
\end{equation}
The expression for tangential pressure can be found following the point
seven of the HJR method.

Thus, the system of surface equations for this model is
\begin{equation}
\label{se1go}\stackrel{.}{A}=F(\Omega -1),
\end{equation}
\begin{equation}
\label{se2go}\stackrel{.}{F}=\frac 1A\left[ 2L+F(1-F)(\Omega -1)\right] ,
\end{equation}
\begin{equation}
\label{se3go}\stackrel{.}{\Omega }=-\frac{\stackrel{.}{F}}F\Omega +\frac{%
\stackrel{.}{K}}K\frac{(1-2K)}{(1-K)}\Omega +\frac{4L\Omega ^2}{3\widetilde{%
\alpha }A^3(2\Omega -1)(1-K)}+\Omega (1-\Omega )\Lambda ,
\end{equation}
and
\[
\stackrel{.}{y}\left[ \frac{2y^4\xi ^2}{2\Omega -1}\left( 4\tau -\frac{{\cal %
C}}\xi \right) -\rho _c\frac{{\cal C}}\xi \left( \frac{K\left( 1-K\right) }{%
K_o}\right) \left( \frac{2\Omega -1}\Omega \right) \right] =\xi ^2y^5\left[
\frac{4\tau \stackrel{.}{\Omega }}{(2\Omega -1)^2}-\frac{2\sqrt{F}}{\sqrt{%
2\Omega -1}}\right] -
\]
\begin{equation}
\label{se4go}\frac{{\cal C}}\xi \left[ \frac{2y^4\xi ^2}{2\Omega -1}+\rho _c%
\frac{K\left( 1-K\right) }{K_o}\left( \frac{2\Omega -1}\Omega \right)
\right] \left( y_1F(2\Omega -1)+y\Phi \right) ,
\end{equation}
where
\begin{equation}
\Lambda =\frac{3\alpha K}{2K_o}A(1-K)\left( \frac{3\Omega -1}\Omega \right) -%
\frac{3+F}{2A}+\frac{2F\Omega }{A(1-K)}(\Psi -K),
\end{equation}
\[
\Psi =\frac 3{10K_o}\lambda \alpha A^2K^2\left( 1-G\right) ^n
\]
\begin{equation}
+\frac{A^2}{2F}\left[ \frac{3\alpha K}{2K_o}\lambda ^2F^2\left( 1-G\right)
^{2n}-\frac{2n\lambda G}{A^2}F^2\left( 1-G\right) ^{n-1}+\frac \alpha 2%
\left( 1-\frac{3K}5\right) \frac{K\left( 1-K\right) }{K_o}\right] ,
\end{equation}
and $\Phi $ is defined in (\ref{fi}).

We have integrated the system of surface equations for this model with the
initial data $A(0)=10; \Omega(0)=1$, and with $n=1; \lambda=1/3; K(0)=0.999$%
. Also, the inital mass is $1.3 M_\odot$. This corresponds to a star with an
initial radius of $19.201$ meters, a central density equal to $2.17 \times
10^{14} g \, cm^{-3}$ and a surface density equal to $2.17 \times 10^{11} g
\, cm^{-3}$. For the luminosity we have taken a gaussian pulse with center
at $u=200$ and width equal to $20$ (in dimenssionless $u$-units), which is
equivalent to a width of $0.13 ms$.

In the case $\tau=0$, the collapse spans over $u \sim 400 \, (2.5 ms)$. As $%
\tau$ increases, that time also increases and may be as large as $32 ms$ for
$\tau=900 \, (5.7 ms)$.

In figure 6 we plot the ratio of the total radiated mass divided by the
total radiated mass for $\tau = 0$, as function of $\tau$, for different
radiated mass in the $\tau=0$ case.

The sensitivity of this ratio to different values of $\tau$ is clearly
exhibited. This fact is also present in the Tolman VI model discussed in the
previous section (fig.5). However in this model non linear effects seem to
be stronger than in the precedent one and the ratio may be larger or smaller
than one, depending on the total radiated mass.

The evolution of the radius is exhibited in figures 7 and 8 for different
total radiated mass ($1\%$ and $0.001\%$ of the total mass) and different $%
\tau$'s. The dependence of the final value of the radius on $\tau$, is
linear for low emission ($0.001\%$) whereas it is not for the stronger case (%
$1\%$).

The ratio $M_{final}/A_{final}$ behaves essentially as the other surface
variables (fig.9). The resulting object is more compact for larger
luminosities, however as $\tau$ increases, the final configuration is less
compact and it may happen that for sufficiently large $\tau$'s, the object
may be at the end of it's evolution less compact than in the $\tau=0$ case.

Finally, figures 10 and 11 show the evolution of luminosity profiles for
different radiated mass and different $\tau$'s. As in all known models,
larger $\tau$'s means longer emissions and more flattened pulses. This
effect is sharper for stronger emissions.

\section{Conclusions}

It has been the purpose of this paper to exhibit the relevance of
thermal relaxation time in the problem of collapse, by means of two models
of radiating spheres. It is important to emphasize that this
relaxation time is systematically neglected in collapse calculations,
where processes may occur on time scales which may be of the order of
magnitude of (or at least not much larger than) relaxation time,
leading thereby to incorrect conclusions.

One of the models (Tolman VI) is suitable for describing the
evolution of the core
at the earliest stages of a supernova explosion.

The second one (G-M) is more adapted to describe the Kelvin-Helmoltz phase
in the birth of a neutron star.

However, due to time restrictions in numerical calculation, in this model
(G-M), we have integrated the surface equation over a period of about tens
of milliseconds, instead of tens of seconds, which is the typical time of
the Kelvin-Helmholtz phase in neutron star formation. Consequently all
times have to be scaled by the same factor.

Preliminary results show that in what the relaxation time concerns, its
influence on the evolution of the object is not qualitatively changed by
this ``scaling''.

In the Tolman VI model, the point to emphasize is the bifurcation introduced
by changes in $\tau$ of the order $10^{-6}$ s. Such values are within the
range of possible values of $\tau$, and therefore the study of the evolution
of such system seems to require a good account of pre-relaxation processes.

In the second model the dependence of the final configuration on $\tau$ may
be non-linear and is affected by the total emission.

At any rate the final configuration is clearly $\tau$-dependent.

It is worth mentioning that the hydrostatic time scale for the
second model is of the order of $0.29 ms$, which clearly indicates that in
these calculations the hydrostatic approximation is not a very good one.
This is an additional argument to use the HJR-formalism, which in some sense
may be envisaged as a ``correction'' to hydrostatic approximation (see point
2 of the algorithm in section 2).

We would like to conclude with the following comment: In a collapse
calculation, neutrino transport plays an important role, the reason
to overlook this issue here, resides in the fact that we are not
concerned with the problem of modeling gravitational collapse, but
with the influence of thermal relaxation time in the outcome of
evolution. Conduction associated with trapped neutrinos is certainly
of greatest relevance, and it should be very interesting to find out
the role played by relaxation time of that process. This however is
out of the scope of this paper.

\section*{Acknowledgments}

This work was partially supported by the Spanish Ministry of Education under
Grant No. PB94-0718

\newpage

\section*{Figure captions}

\begin{description}
\item[Figure 1.-]  Luminosity profile as a function of time. In all figures
the values of the relaxation time are given in miliseconds.

\item[Figure 2.-]  Evolution of the temperature in the surface for different
values of $\tau $.

\item[Figure 3.-]  Evolution of $M/A$ for different values of $\tau $. Note
the bifurcation between $\tau =0.0032$ ms and $\tau =0.0038$ ms.

\item[Figure 4.-]  Radius as a function of time. The bounce is only possible
for $\tau <0.0035$ ms.

\item[Figure 5.-]  Ratio between total radiated mass and total radiated mass
for $\tau =0$ as a funtion of the relaxation time. Dashed and solid lines
correspond to TOL04 and TOL06 models respectively. TOL04 model fails for $%
\tau$ larger than the corresponding to point marked A.

\item[Figure 6.-]  Same as fig. 5 for G-M model. Curves are labeled with the
percentage of radiated mass in the case $\tau=0$.

\item[Figure 7.-]  Variation in the radius for $M_r=10^{-5}$ (0.001\%)

\item[Figure 8.-]  Same as fig. 7 for $M_r=0.01$ (1\%). Note that in the low
emission case (fig. 7) the variation is given in cm, while in this case $%
\Delta A$ is given in meters.

\item[Figure 9.-]  Ratio between final value of $M/A$ and $M/A$ for $\tau=0$
as a function of $\tau$. The labels in each curve means the same as in fig.
6.

\item[Figure 10.-]  Luminosity profile for different values of $\tau$. The
radiated mass in the case $\tau=0$ is a 0.001\% of the initial mass.

\item[Figure 11.-]  Same as fig. 10. The radiated mass for $\tau=0$ is a 1\%
of the initial mass.
\end{description}

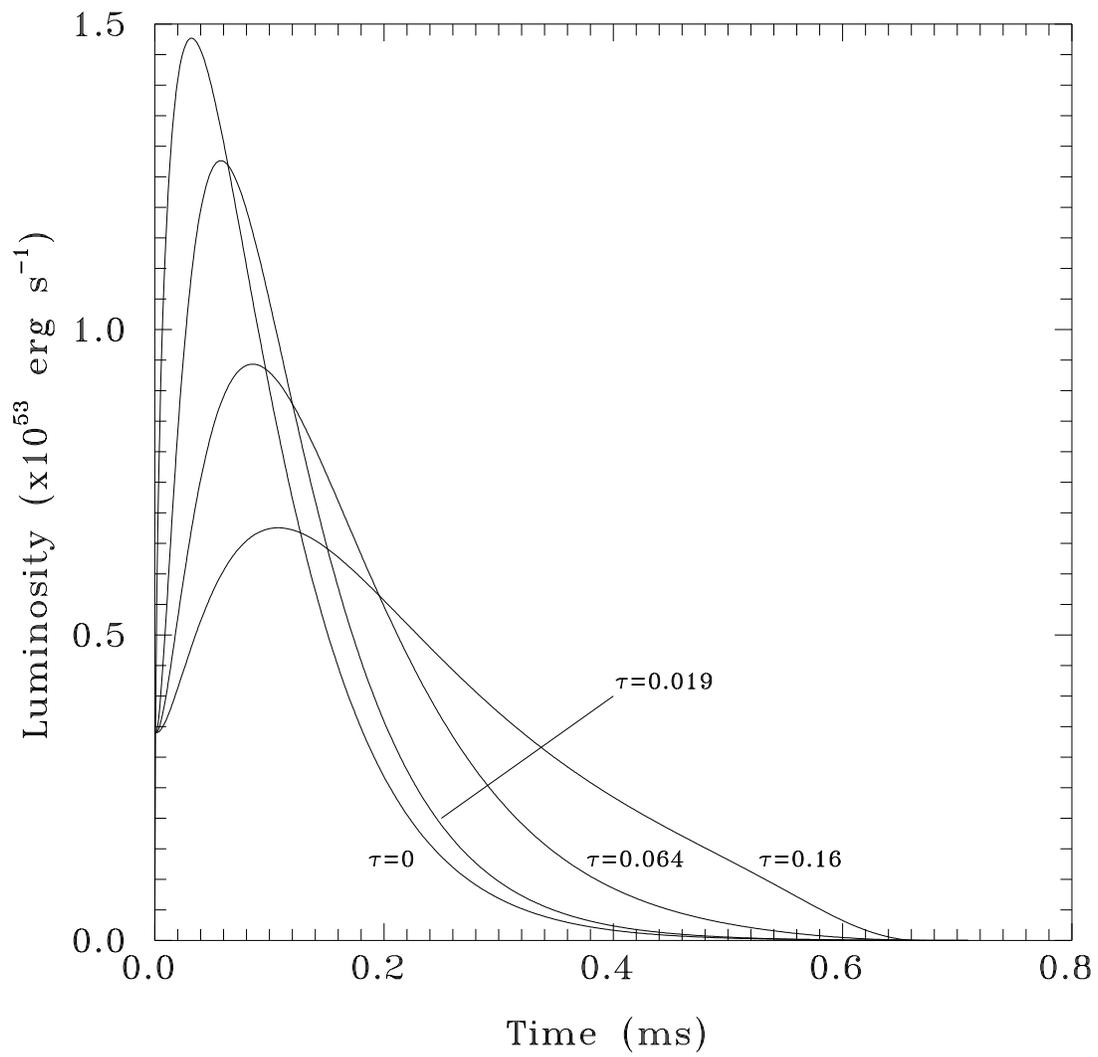
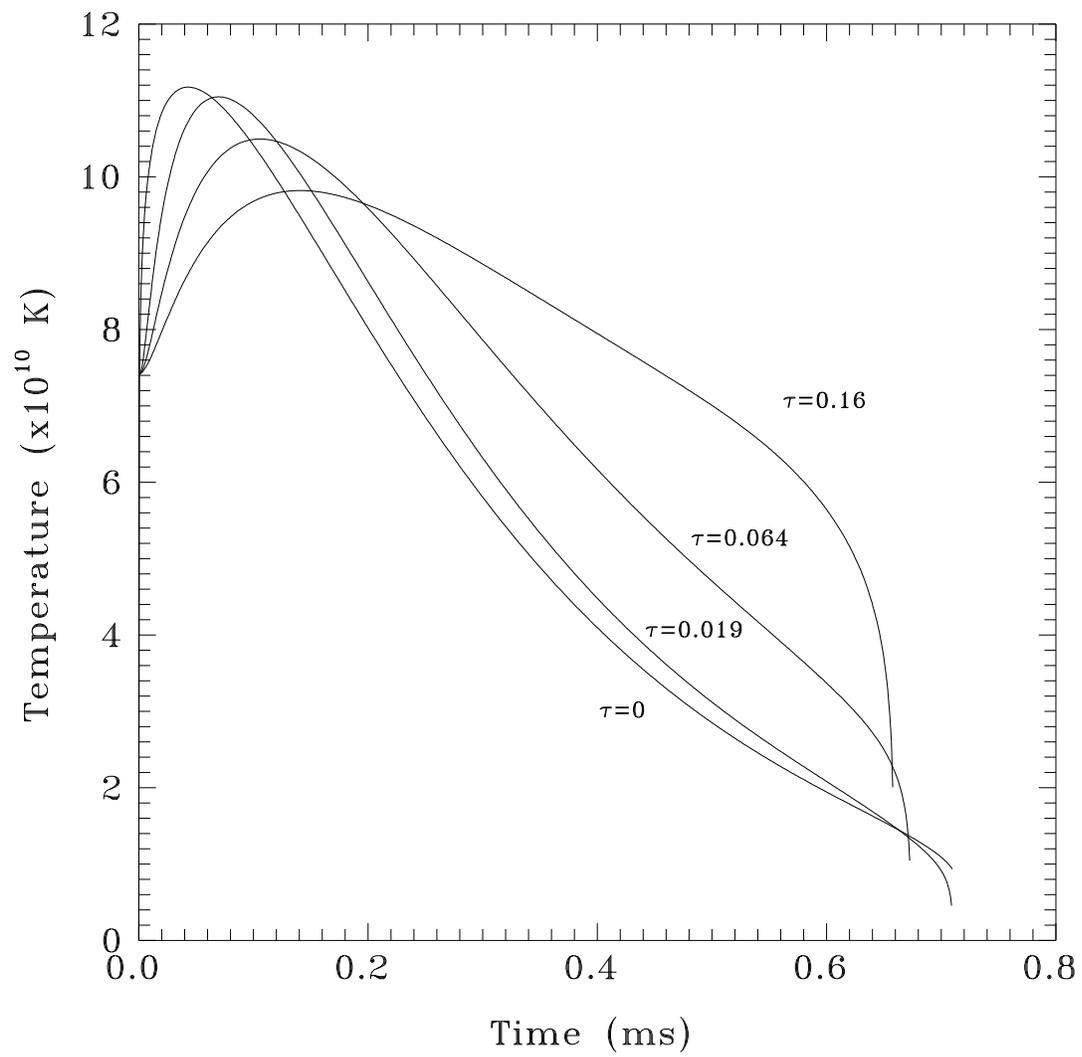
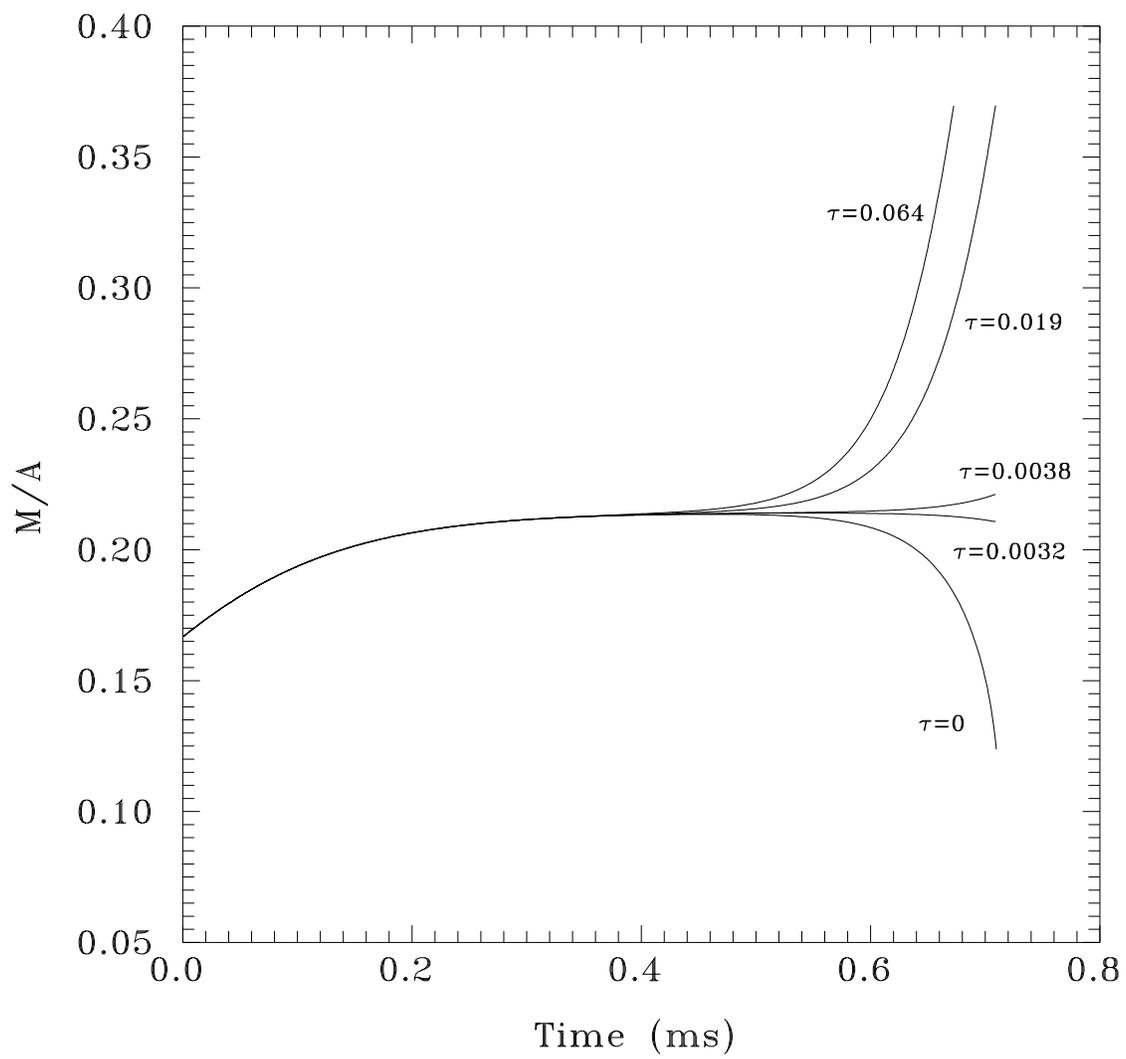
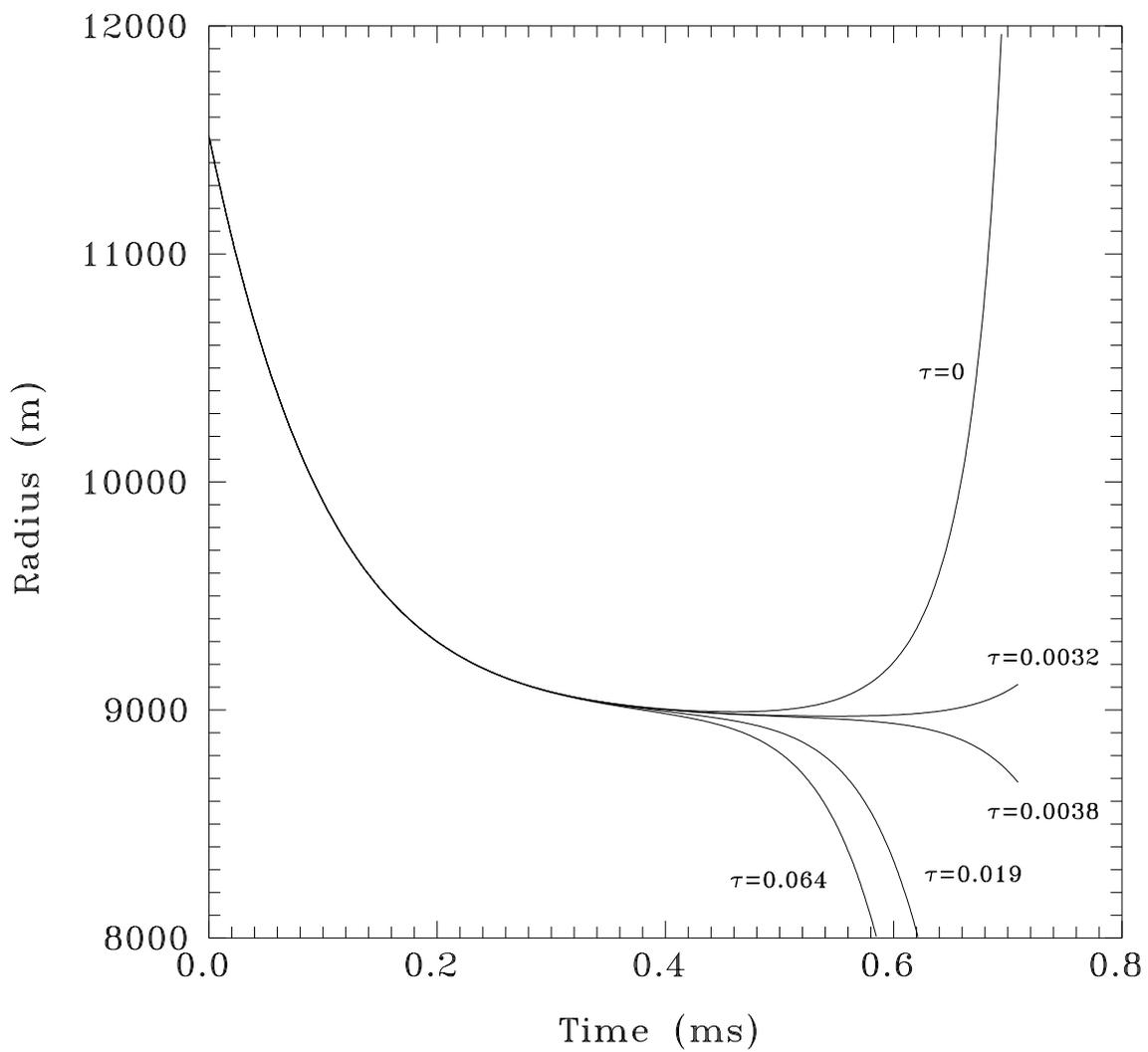
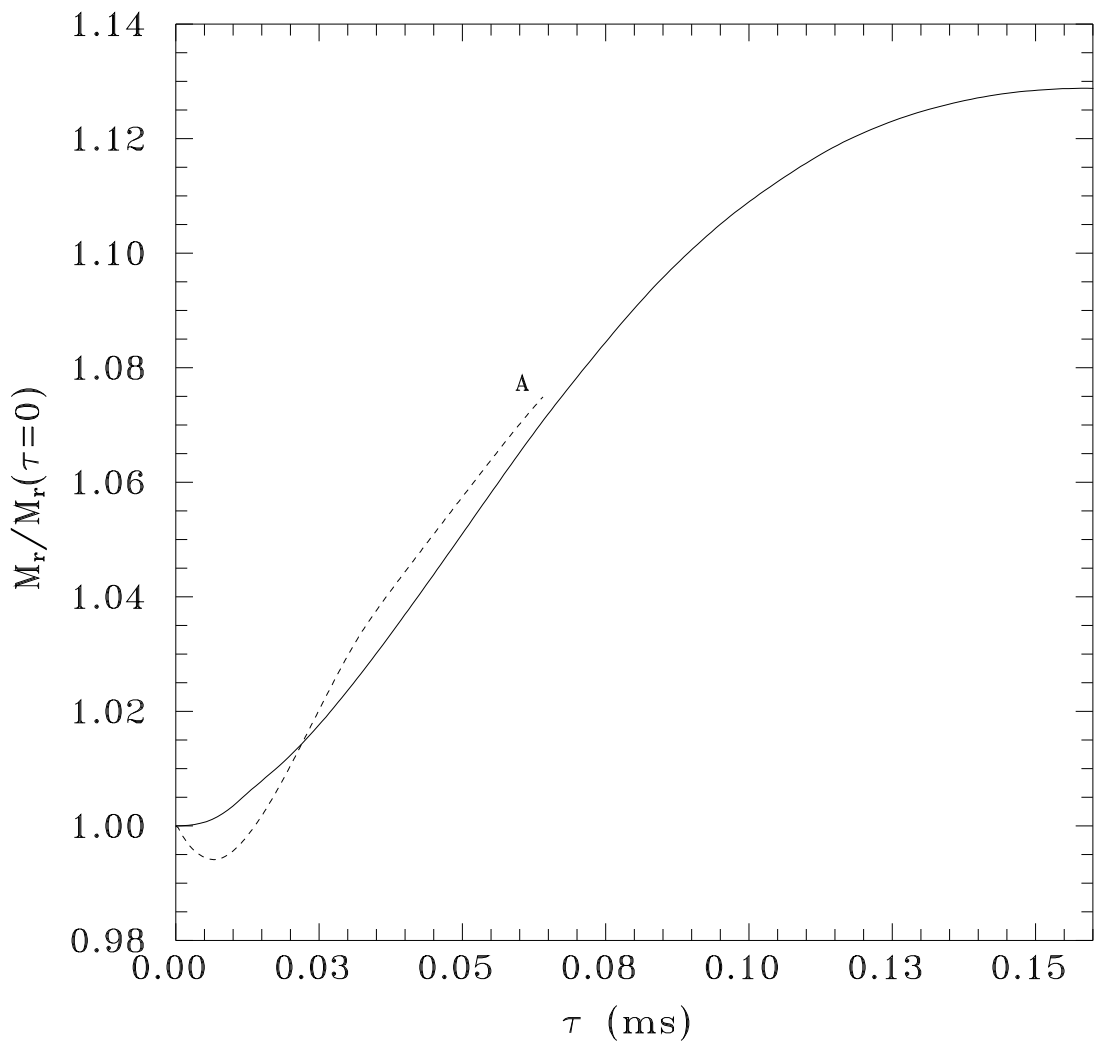
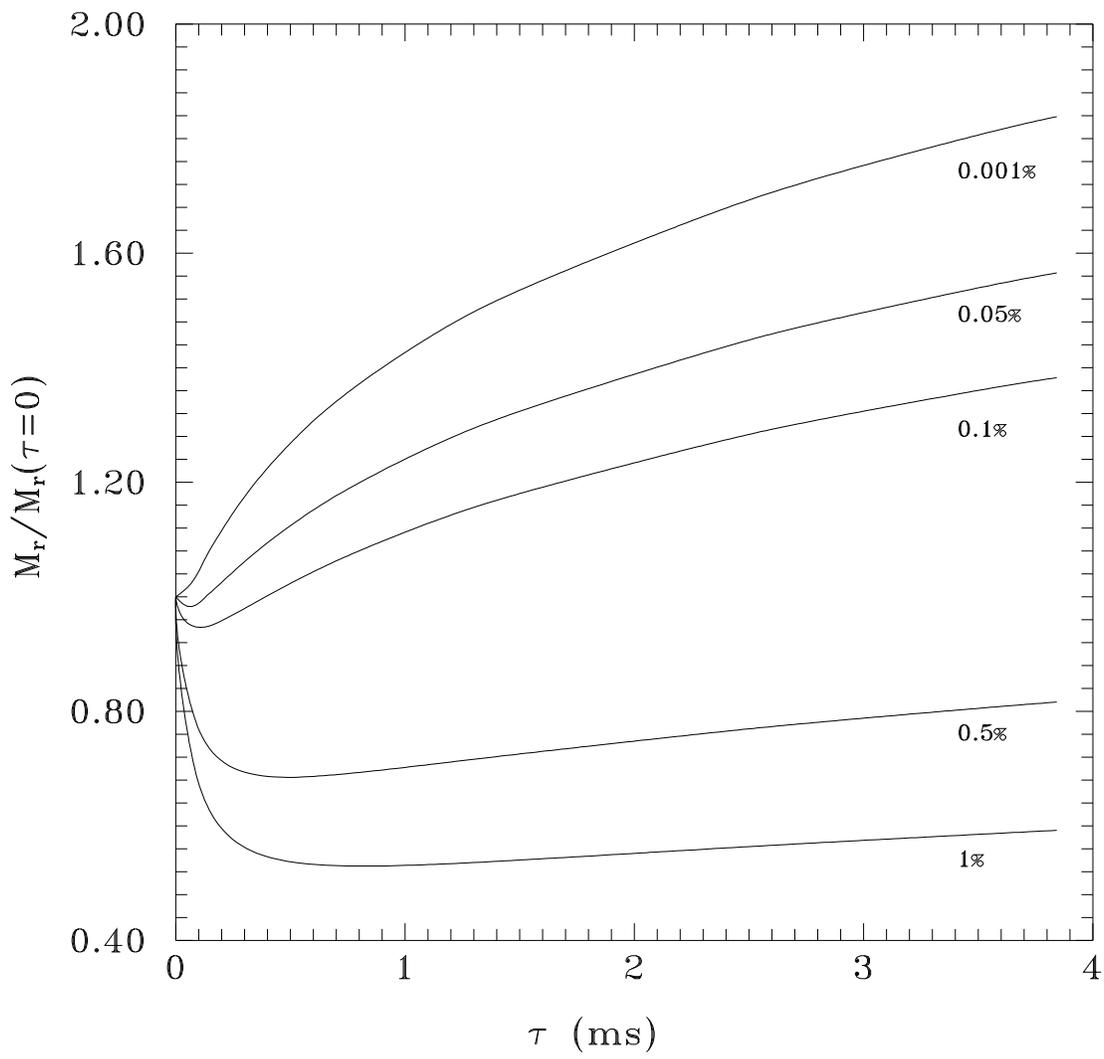
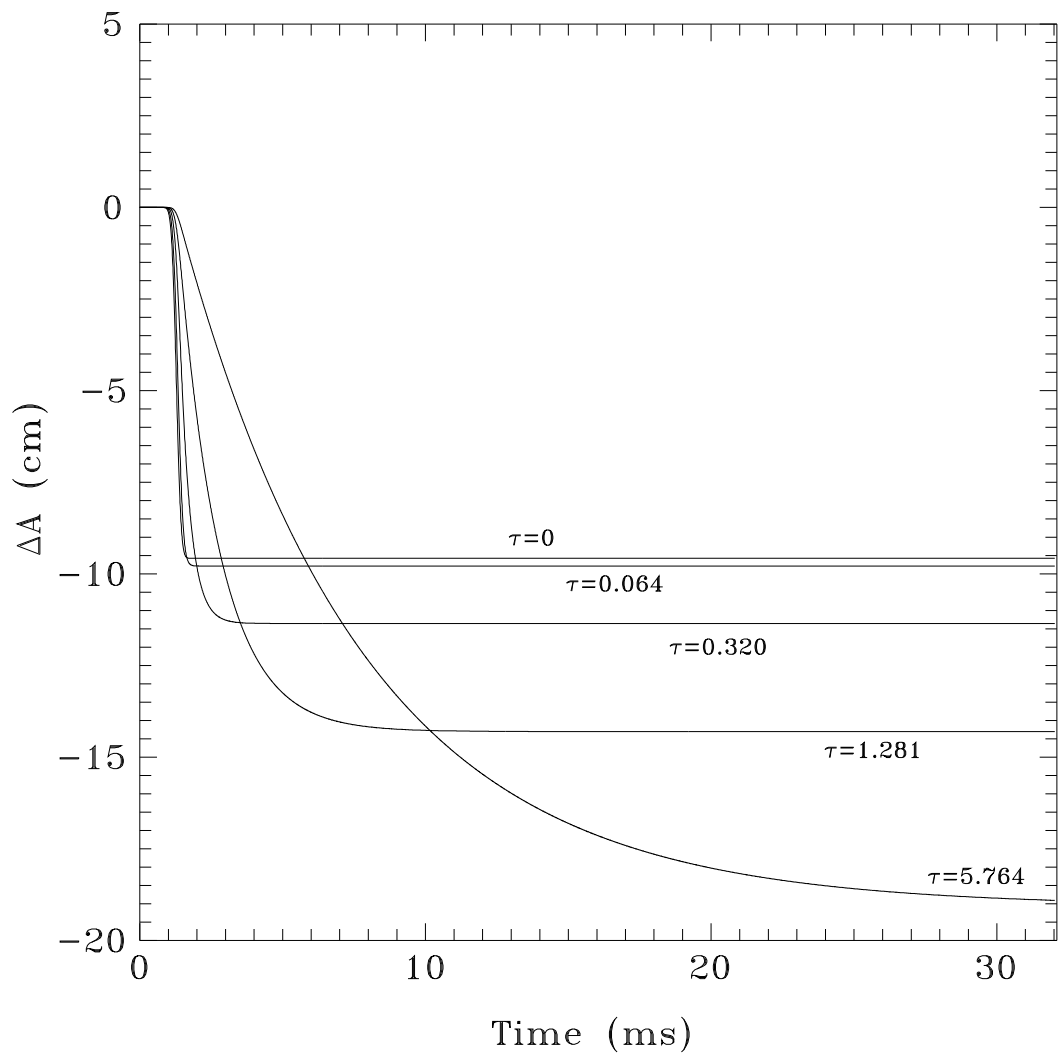
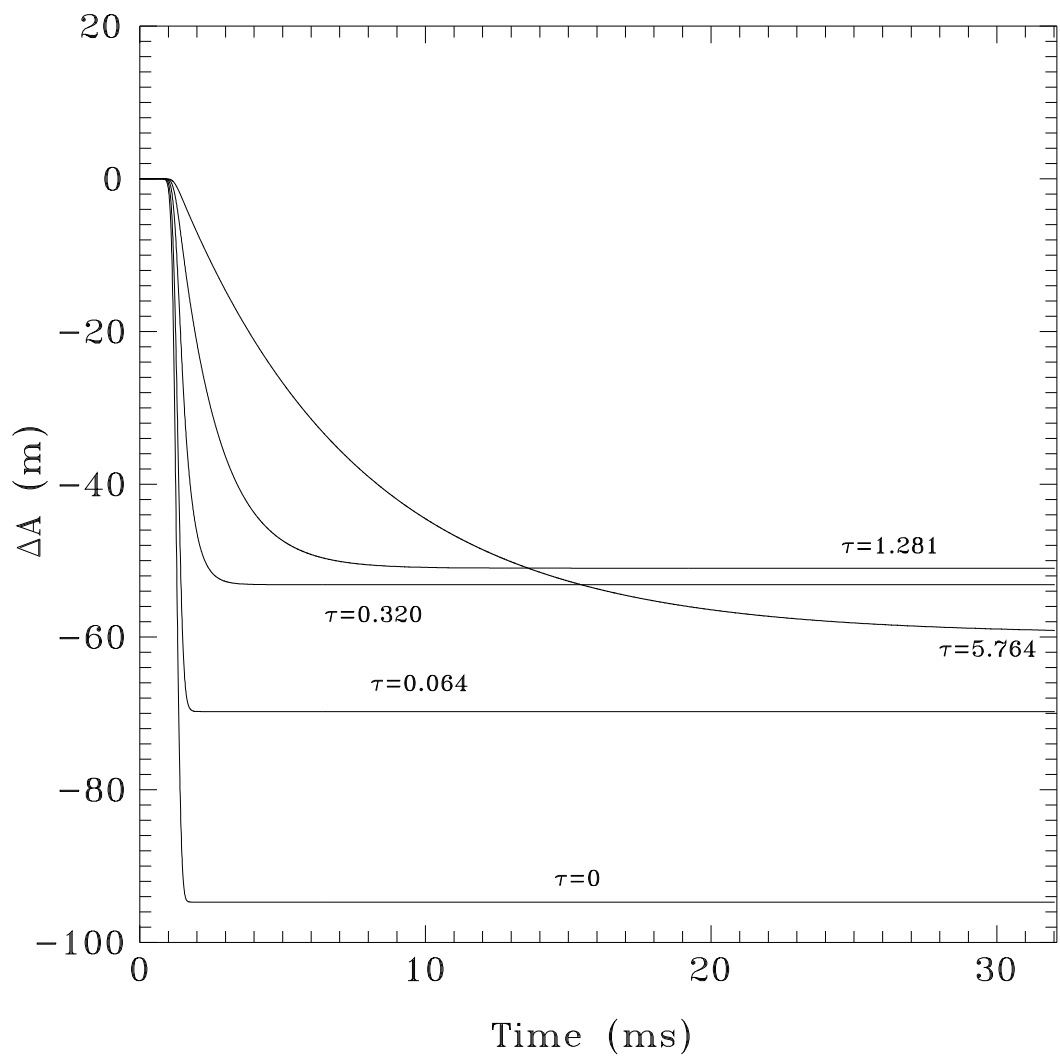
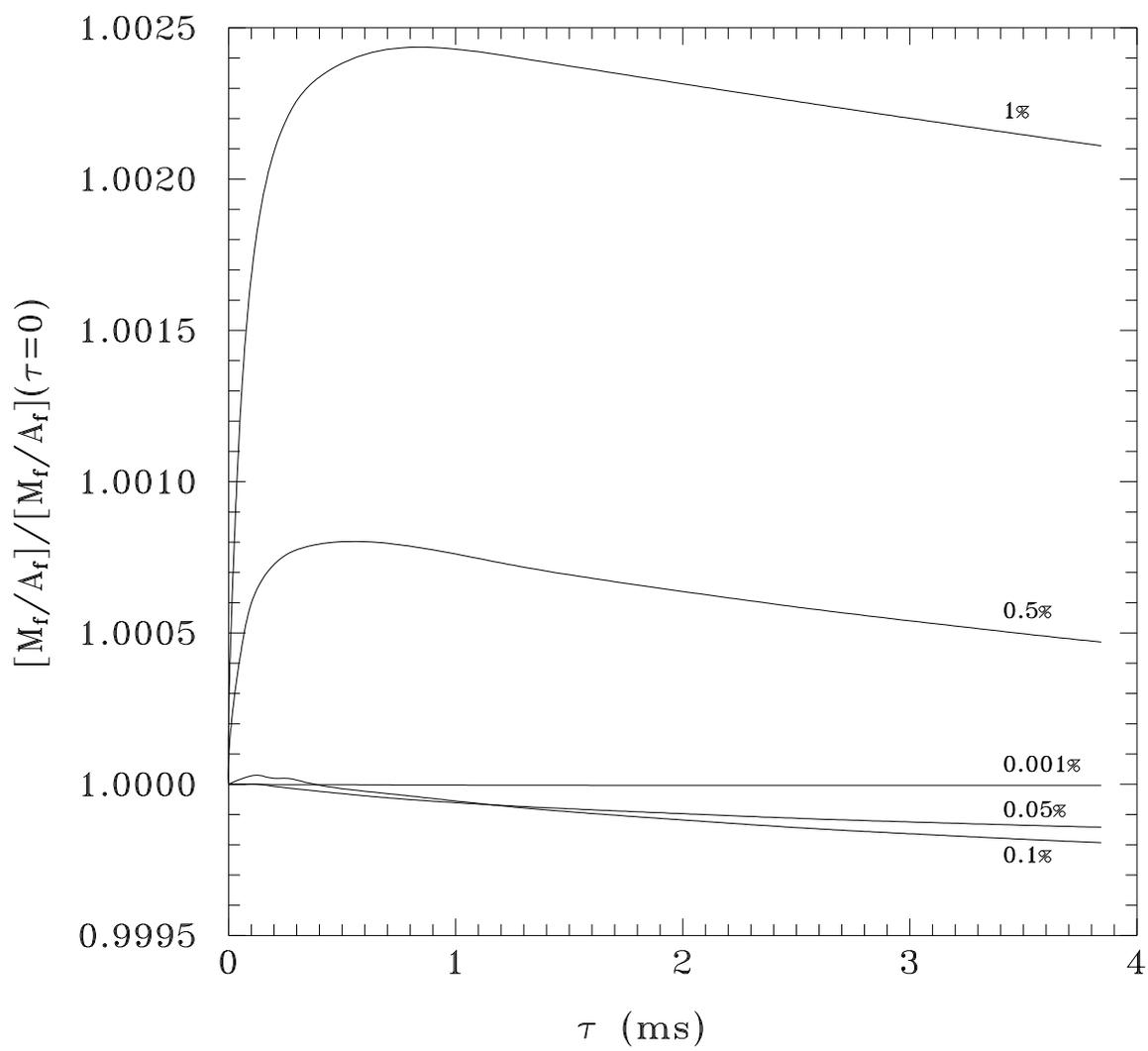
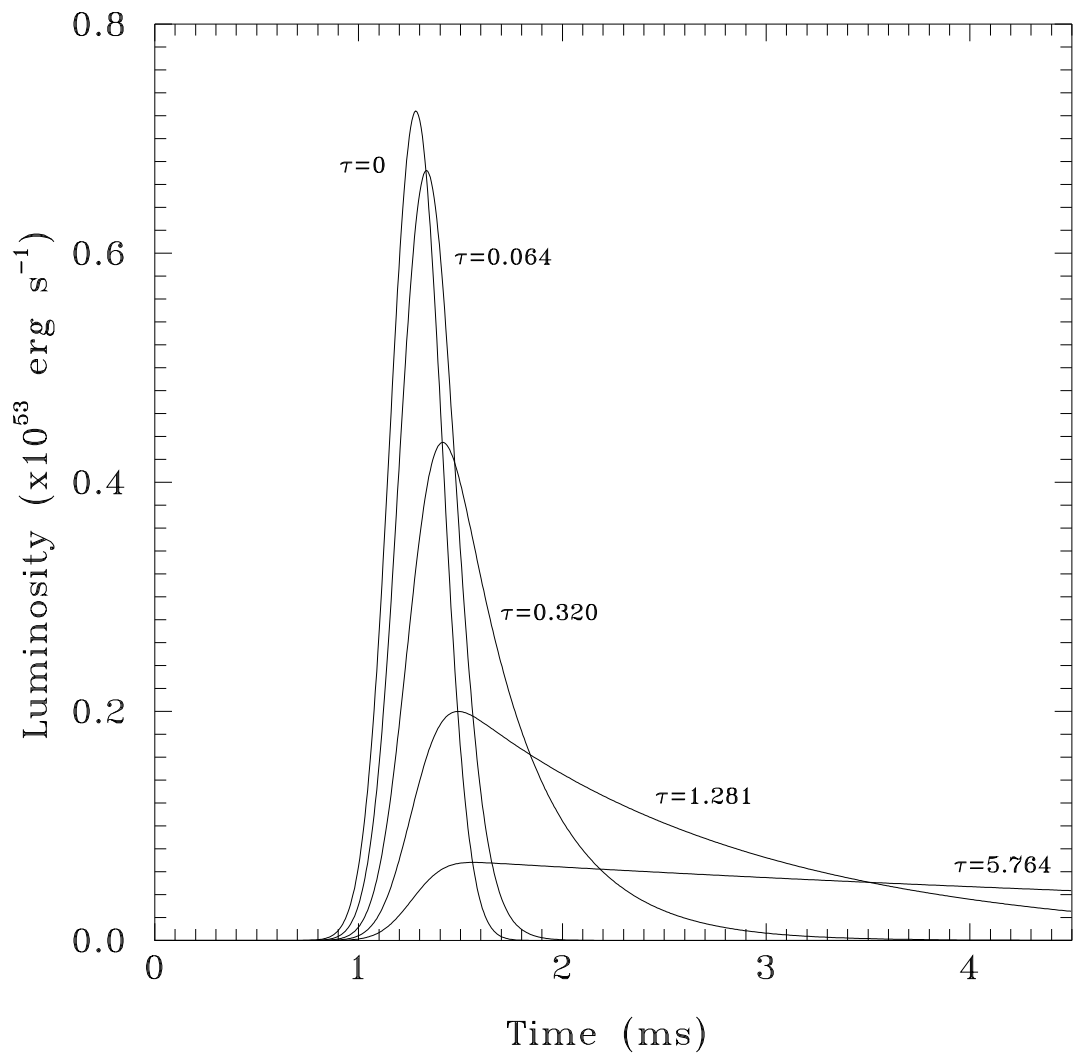
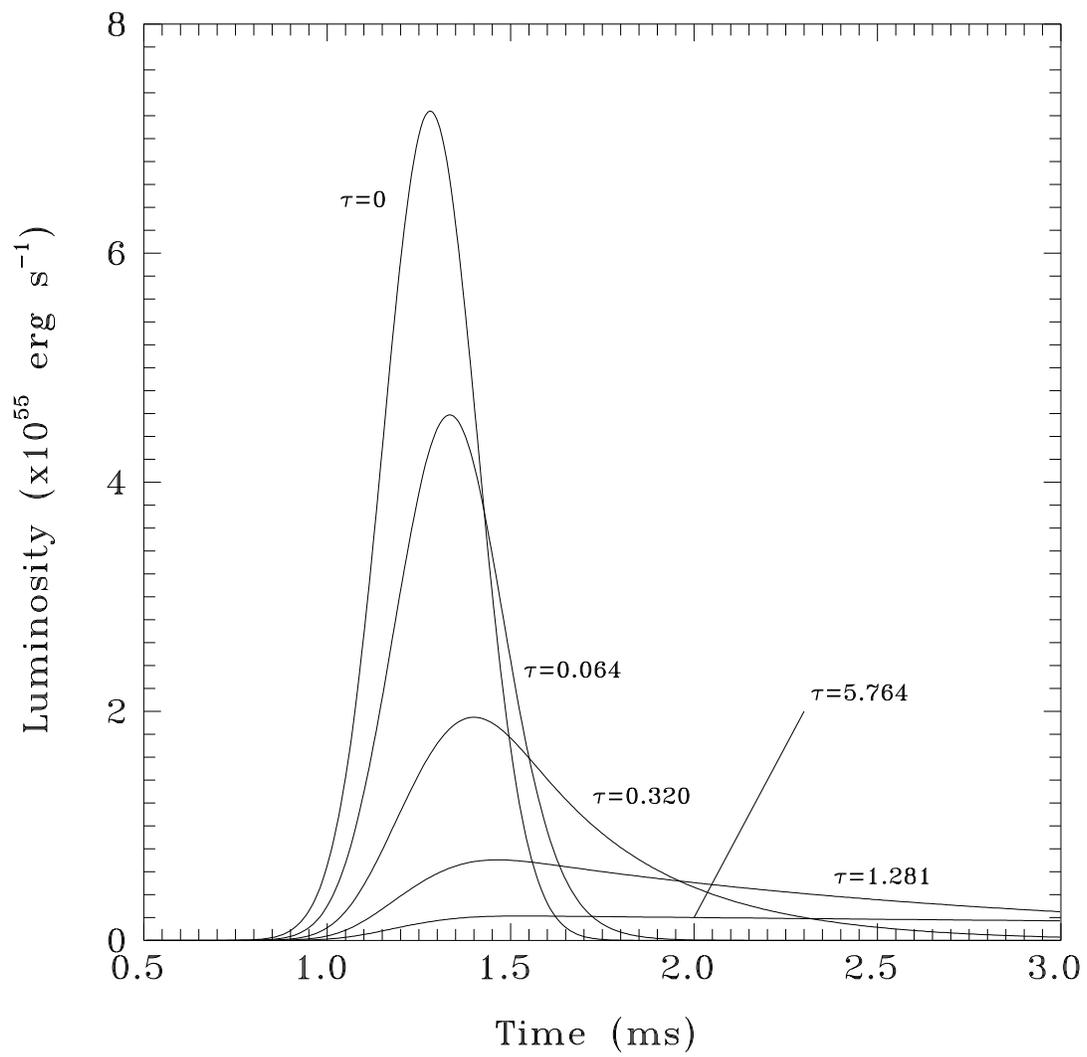
\end{document}